\documentclass[aps,prb,reprint,twocolumn,showpacs,floatfix,superscriptaddress]{revtex4}
\usepackage{graphicx}
\usepackage{amssymb}
\usepackage{amsmath}
\usepackage{lmodern}
\usepackage{color}

\begin{document}

\title{Raman scattering and electrical resistance of highly disordered graphene}

\author{I. Shlimak}
\email{Issai.Shlimak@biu.ac.il}
\affiliation{Jack and Pearl Resnick Institute, Department of Physics, Bar-Ilan University, Ramat-Gan 52900, Israel}

\author{A.Haran}
\affiliation{Faculty of Engineering, Bar-Ilan University, Ramat-Gan 52900, Israel}

\author{E. Zion}
\affiliation{Department of Physics and Institute of Nanotechnology and Advanced Materials, Bar-Ilan University, Ramat-Gan 52900, Israel}

\author{T. Havdala}
\affiliation{Department of Physics and Institute of Nanotechnology and Advanced Materials, Bar-Ilan University, Ramat-Gan 52900, Israel}

\author{Yu. Kaganovskii}
\affiliation{Jack and Pearl Resnick Institute, Department of Physics, Bar-Ilan University, Ramat-Gan 52900, Israel}

\author{A. V. Butenko}
\affiliation{Department of Physics and Institute of Nanotechnology and Advanced Materials, Bar-Ilan University, Ramat-Gan 52900, Israel}

\author{L. Wolfson}
\affiliation{Jack and Pearl Resnick Institute, Department of Physics, Bar-Ilan University, Ramat-Gan 52900, Israel}

\author{V. Richter}
\affiliation{Solid State Institute and Physics Department, Technion-Israel Institute of Technology, Haifa 32000, Israel}

\author{D. Naveh}
\affiliation{Faculty of Engineering, Bar-Ilan University, Ramat-Gan 52900, Israel}

\author{A. Sharoni}
\affiliation{Department of Physics and Institute of Nanotechnology and Advanced Materials, Bar-Ilan University, Ramat-Gan 52900, Israel}

\author{E. Kogan}
\affiliation{Jack and Pearl Resnick Institute, Department of Physics, Bar-Ilan University, Ramat-Gan 52900, Israel}

\author{M. Kaveh}
\affiliation{Jack and Pearl Resnick Institute, Department of Physics, Bar-Ilan University, Ramat-Gan 52900,
Israel}

\date{\today}

\begin{abstract}
Raman scattering (RS) spectra and current-voltage characteristics at room temperature were measured in six series of small samples fabricated by means of electron-beam lithography on the surface of a large size (5x5 mm) industrial monolayer graphene film. Samples were irradiated by different doses of C${}^+$ ion beam up to $10^{15}$ cm${}^{-2}$. It was observed that at the utmost degree of disorder, the Raman spectra lines disappear which is accompanied by the exponential increase of resistance and change in the current-voltage characteristics. These effects are explained by suggestion that highly disordered graphene film ceases to be a continuous and splits into separate fragments. The relationship between structure (intensity of RS lines) and sample resistance is defined. It is shown that the maximal resistance of the continuous film is of order of reciprocal value of the minimal graphene conductivity
$\pi h/4e^2\approx 20$ kOhm.
\end{abstract}

\pacs{73.22.Pr}

\maketitle

\section{Introduction}

In recent years, disordered graphene attracts the attention of many researchers \cite{katsnelson,roche,warner}. Mainly, this is due to the possibility of obtaining a high-resistance state of graphene films, which is of interest for application in electronics. In the experiment, the disorder is achieved in various ways: by oxidation \cite{liu}, hydrogenation \cite{elias}, chemical doping \cite{liu2}, irradiation by different ions with different energies \cite{saito,buchowicz,guo,wang,ferrari}. In this work the ion bombardment technique is used to gradually induce disorder in graphene. To probe the evolution of disorder, the Raman spectroscopy and resistance measurements are  used.

The main new results of this work consist in (i) observation of the utmost degree of disorder induced by ion irradiation, when graphene, due to high density of defects, is no longer continuous film but split into separate fragments. This leads to complete disappearance of the Raman scattering spectra (RS) and change in the mechanism of electrical conductivity; (ii) determination of a correlation between intensity of RS lines and sample resistance in disordered graphene films: transition from the low-defect density regime to the high-defect density regime in RS occurs at the resistance equal to reciprocal value of the minimal graphene conductivity $\pi h/4e^2\approx 20$ kOhm.

\section{Samples}

The initial specimens were purchased in "Graphenea" company. Monolayer graphene was produced by CVD on copper catalyst and transferred to a 300 nm SiO${}_2$/Si substrate using wet transfer process. The specimen size was $5\times 5$ mm.  Graphene film of such a large size was not a monocrystalline. It looks like a polycrystalline film with the average size of microcrystals about few microns (Fig. 1).

On one of these specimens, gold electrical contacts were deposited directly on the graphene surface through a metallic mask. This sample is marked as "0". On the other 5x5 mm specimen, many samples of small size ($200\times 200$ $\mu$m) as well as electrical contacts (5 nm Ti and 45 nm Pd) thereto were made using electron beam lithography (EBL). Optical observation and electrical testing after EBL showed that some of small samples are damaged, the total yield was about 65$\%$. This reduced yield can be explained by the influence of the triple lift-off process in the case of large size polycrystalline graphene film. One can assume that the boundary between single crystals is a weakness which can be damaged during the lift-off process. For further measurements, only intact samples were selected.

\begin{figure}[h]
\includegraphics[width= \columnwidth]{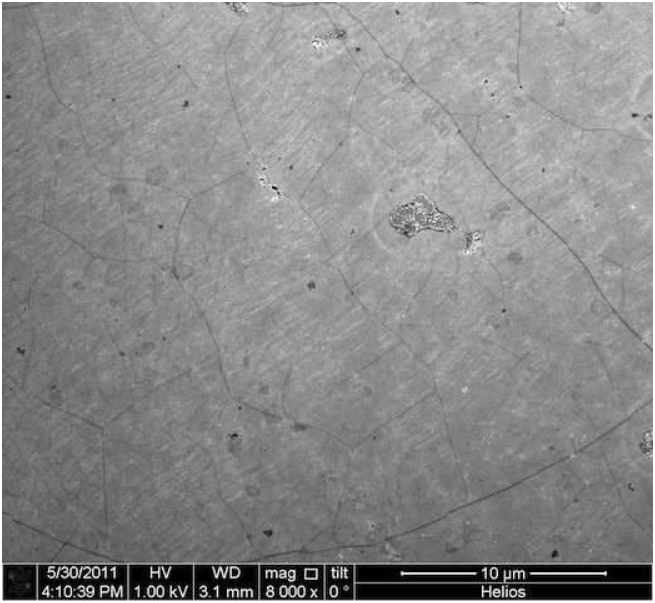}
\caption{\label{fig:bands}  Microphotograph of the graphene monolayer specimen }
\end{figure}

All small samples on the surface of the common 5x5 mm specimen were grouped into 6 groups. The first group was not irradiated. Samples from this group have a mark "1".  Groups "2-6" were irradiated by C${}^+$ ions with energy 35 keV at five different doses: $5\times 10^{13}$, $1\times 10^{14}$, $2\times 10^{14}$, $4\times 10^{14}$ and $1\times 10^{15}$ cm${}^{-2}$ correspondingly. Samples from these groups have marks from 2 to 6. Below, we report the results of measurements of the RS spectra and two-probes electrical resistance in these samples at room temperature

\section{Raman scattering}

In the RS spectra measurements, excitation was realized by the laser beam with $\lambda = 532$ nm and power less than 2 mW to avoid heating and film destruction. Reproducibility was verified by repeated measurements in the same point and in different samples of the same group. Fig.
2a shows the RS in initial sample "0" and sample 1 which was not irradiated. Initial sample "0" has a typical RS structure for monolayer graphene film \cite{saito}: there are three main lines in the spectrum: a weak D-line at 1350 cm${}^{-1}$ related to the inter-valley double resonant process in graphene with defects (edge, vacancy, etc.), a 2D-line at 2700 cm${}^{-1}$, related to an inter-valley two phonon mode, characteristic for the perfect crystalline honeycomb structure, and a graphite-like G-line at 1600 cm${}^{-1}$ which is common for different carbon-based films. Usually, the ratio $I_D/I_G$ between D and G lines is used as a measure of disorder in graphene films.
The ratio $I_D/I_G$ for the sample "0" is 0.15. This means that the initial film has a not bad quality though not being a perfect. It looks natural for the large size polycrystalline films. The perfect films with RS without D-line are usually obtained in the case of graphene flakes of small size about 10 $\mu$m or less.

Measurements of RS spectra in sample 1 showed that the ratio $I_D/I_G$ increases from 0.15 up to 1.8 (Fig. 2a). It means that EBL introduces disorder itself, even without ion irradiation. Usually, in the case of small flakes, EBL does not lead to an increase of disorder. The damaging effect of EBL in our case could be explained by the presence of border regions in the polycrystalline film. The boundary regions may have a reduced adhesion to the SiO${}_2$ substrate and therefore could be damaged by lift-off process during EBL.

\begin{figure}[h!]
\includegraphics[width=\columnwidth]{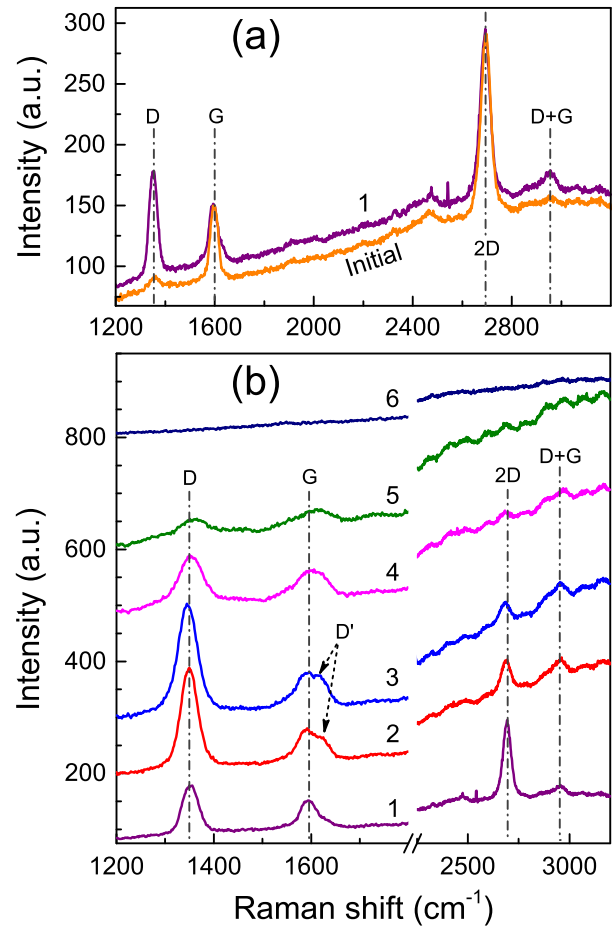}
\caption{  (Color online) RS spectra: (a) - initial sample "0" and sample 1 (after EBL, non-irradiated); (b) - samples 1-6.  $\Phi$, cm${}^{-2}$: 1- 0, 2 - $0.5\times 10^{14}$, 3 - $1\times 10^{14}$, 4 - $2\times 10^{14}$, 5 - $4\times 10^{14}$, 6 - $1\times 10^{15}$. The lines are shifted for clarity}
\end{figure}

Irradiation leads to further increase of disorder. Fig. 2b shows the transformation of the RS spectra in samples 2-6 with increasing irradiation dose $\Phi$. Up to $\Phi = 1\times 10^{14}$ cm${}^{-2}$, the amplitude of the "defect" D-line increases, while the "perfect" 2D-line quickly disappears; new "defect" lines appear: D'-line (1620 cm${}^{-1}$) and (D+G)-line (2950 cm${}^{-1}$). As to G-line, it remains approximately constant and only broadened because of appearance of the nearly located D'-line. So, intensity of D-line increases up to $I_D/I_G\approx 3.2$ at $\Phi = 10^{14}$ cm${}^{-2}$.
Further increase of the irradiation leads to decrease and broadening of D-line, so the ratio $I_D/I_G$ decreases. On the latter stages, G-line also broadened, decreases and eventually, all RS structure disappears at $\Phi = 1\times 10^{15}$ cm${}^{-2}$ (sample 6). We are not aware about observation of complete disappearance of all RS lines in disordered graphene. Usually dependence of $I_D/I_G$ as a function of disorder displays two different behaviors. In the regime of "low-defect-density", $I_D/I_G$ increases with increase of the irradiation dose $\Phi$. In the regime of "high-defect-density", $I_D/I_G$ decreases with further increase of $\Phi$, which is explained by amorphization of the graphen structure attenuating all Raman peaks. However, complete disappearance of all RS lines cannot be explained by amorphization, because G-line is observed even in amorphous carbon films \cite{ferrari2}. We assume that at the maximal level of disorder achieved in this experiment, the graphene film ceases to be a continuous and splits into separate spots of small size (quantum dots). Small size of the quantum dots makes it impossible to form phonons responsible for the structural line of RS.

Degree of disorder can be characterized by the concentration of defects $N_D$ (cm${}^{-2}$) or by the mean distance between defects $L_D =N_D^{-1/2}$. Our irradiation conditions (C${}^+$ ions with energy 35 keV)   were chosen such that the end-of-range damage would be away from the graphene film \cite{buchowicz}. In this case, concentration of induced defects $N_D$ (vacancies, double vacancies, complex scattering centers, edges, etc.) is much less than the dose of irradiation $\Phi$: $N_D = k\Phi$, where $k\ll 1$.

We will use the empirical model which describes the dependence of $I_D/I_G$ vs. $L_D$ in both "low-defect-density" and "high-defect-density" regimes, that has been developed in Ref. \cite{lucchese}.  In this model, a single defect causes modification of two length scale $r_A$ and $r_S$
($r_A > r_S$). Just in the near vicinity of the defect, the area $S = \pi r_S^2$ is structurally disordered, but at $r_S < r < r_A$, the lattice structure is saved, though the proximity to a defect leads to breaking of selection rules
and emission of D-line. The "activated" area responsible for D-line is $A = \pi(r_A^2 - r_S^2)$. In the "low-defect-density" regime, intensity of D-line linearly increases with $N_D$ which means that $I_D/I_G \sim L_D^{-2}$ (the ratio $I_D/I_G$ is used to normalize the intensity of D-line in comparison between different measurements). The maximal value of $I_D$ is achieved when $L_D$ decreases down to $r_A$ ($L_D \approx r_A$). Further decrease of $L_D$ leads to overlap between $A$-and $S$-areas which results in decrease of $I_D/I_G$. The rate equations which describes evolution of the S- and A-regions are
\begin{eqnarray}
\label{s1}
\frac{dS_S}{dN}&=&\pi r_S^2\left(\frac{S_T-S_S}{S_T}\right)\\
\label{s2}
\frac{dS_A}{dN}&=&\pi (r_A^2-r_S^2)\left(\frac{S_T-S_S-S_A}{S_T}\right)-\pi r_S^2\frac{S_A}{S_T}.
\end{eqnarray}
Here  $S_S, S_A$ and $S_T$ denote the structurally disordered, activated and total areas of the film and $N$ is the number of the induced defects. The last term in Eq. (\ref{s2}) describes the overlap of $S$- and $A$-areas, which was  omitted in \cite{lucchese}. With this term, the final equation for the dependence $I_D/I_G(L_D)$ has the form:
\begin{eqnarray}
\label{final}
I_D/I_G&=&C_Ae^{-\pi r_S^2/L_D^2}\left[1-e^{-\pi (r_A^2-r_S^2)/L_D^2}\right]\nonumber\\
&+&C_S\left[1-e^{-\pi r_S^2/L_D^2}\right].
\end{eqnarray}
Actually, Eq. (\ref{final}) could be obtained directly from basic probability theory, taking into account that for the  Poisson distribution of the defect positions the quantity $e^{-\pi r_S^2/L_D^2}$ is just the probability  that there are no defects in the  circle with radius $r_S$.
Thus the multiplier $1-e^{-\pi r_S^2/L_D^2}$ gives the fraction of the structurally-disorder region; similarly the multiplier $e^{-\pi r_S^2/L_D^2}\left[1-e^{-\pi (r_A^2-r_S^2)/L_D^2}\right]$ gives the fraction of the activated region.

\begin{figure}[h!]
\includegraphics[width=.9\columnwidth]{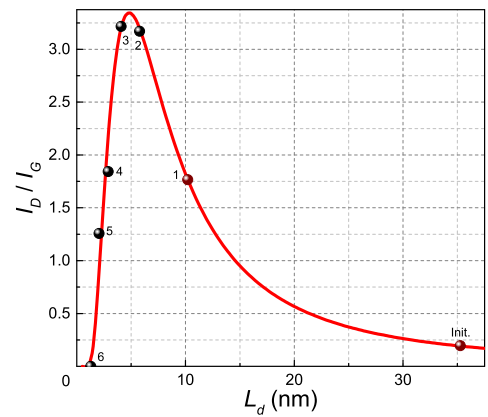}
\caption{Ratio $I_D/I_G$ as a function of the mean distance between defects LD. Solid line represents  Eq.(\ref{final}) with $C_A = 5.4, C_S = 0, r_S= 1.55$ nm, $r_A = 4.1$ nm.}
\end{figure}

Fig. 3 shows the result of fitting the theoretical curve (Eq.(3)) with experimental data. First, the experimental points for irradiated samples 2-6 were plotted.
Then, the curve (Eq.(3)) was  fitted to the experimental points.

The value of $k$ was estimated on the basis of atomic computer simulation \cite{lehtinen}. In that work, the probabilities of various graphene defect production by ion irradiation was calculated by the analytical potential molecular dynamics. The calculations were done  for ions of different masses and  different energies. Interpolating these results to the mass of carbon ion (12 a.u) with energy 35 keV, we obtained the value $k = 6\%$. Similar value of $k$ was used previously in Ref. \cite{buchowicz}.
Thus $L_D$ for our samples was defined as $L_D = (0.06\Phi)^{-1/2}$.

  Finally, the values of $I_D/I_G$ for non-irradiated samples "0" and 1 were placed "by hand" on the curve. This allows to estimate $L_D$ for these samples and, therefore, the density of defects $N_D$ in initial film as $8\times 10^{10}$ cm${}^{-2}$
and in sample 1 as $1\times 10^{12}$ cm${}^{-2}$. As a
result, we can plot the RS data for all samples on the scale $I_D/I_G$ as a function of $N_D$. This dependence is shown in Fig. 4.

\section{Sample resistance}

In Fig. 4, the resistance R at room temperature for all samples is also shown. Resistance of "initial" sample "0", is $R\approx 600$ Ohm, which agrees with the numerous data reported for the resistivity of pristine graphene (in our samples, resistivity is equal to resistance due to square geometry).One can see that increase of $N_D$ leads to the strong continuous increase of $R$ over many orders of magnitude.
Fig. 4 demonstrate also a relationship between structure and resistance of disordered graphene. Maximum value of the ratio $I_D/I_G$ occurs at
$N_D \approx 4\times 10^{12}$ cm${}^{-2}$, when $L_D$ (5 nm) is indeed close to $r_A$ (4.1 nm). In other words, at this $N_D$, the "activated" $A$-area fills all the sample space. Fig. 4 shows also, that the maximum $I_D/I_G$ corresponds to the film resistance $R_{max}\approx 20$ kOhm, which is approximately equal to the reciprocal value of the minimal conductivity in graphene $4e^2/\pi h$ (see e.g. Ref. \onlinecite{ziegler}) or to the resistance quantum $h/e^2\approx 25.8$ kOhm. This allows us to conclude that the graphene film with $R\gg R_{max}$ is not continuous which supports the assumption about fragmentation of the highly disordered graphene film which was made on the basis of disappearance of the RS lines.

\begin{figure}[h!]
\includegraphics[width=.9\columnwidth]{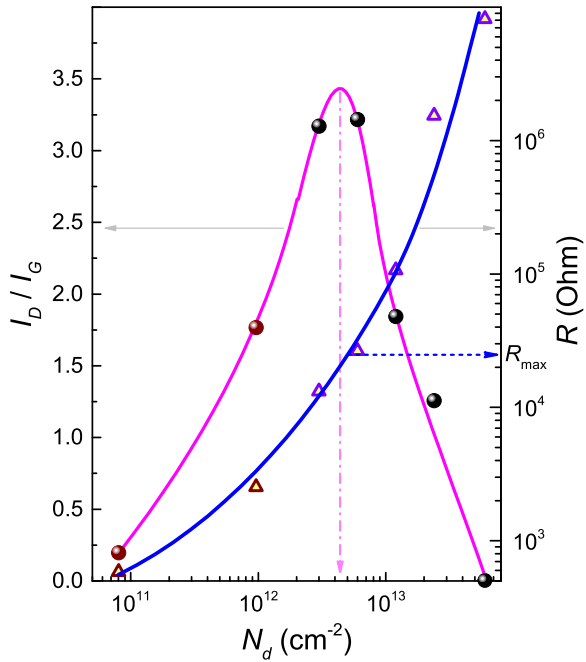}
\caption{The RS line ratio $I_D/I_G$ and the sample resistance $R$ at 300 K plotted as a function of the defect concentration $N_D$. }
\end{figure}

Measurements of the current-voltage characteristics ($I-V$) showed that for weakly disordered films (samples 1 - 4), $I-V$ characteristics is linear (Ohmic regime), while for highly disordered films (samples 5 and 6), $I-V$ is strongly non-linear. Fig. 5 shows the static resistance $R = V/I$ as a function of the current for samples 2 - 6. For samples 5 and 6, $R$ has the maximal
value $R_0$ at $I\to 0$ and falls with increase of $I$ or $V$. The values of $R_0$ are plotted for samples 5 and 6 in Fig. 4.
The change in the character of $I-V$ for highly irradiated samples can be explained by the replacement of usual conductivity along the continuous film by the field-induced electron tunneling between quantum dots.

\begin{figure}[h!]
\includegraphics[width=.9\columnwidth]{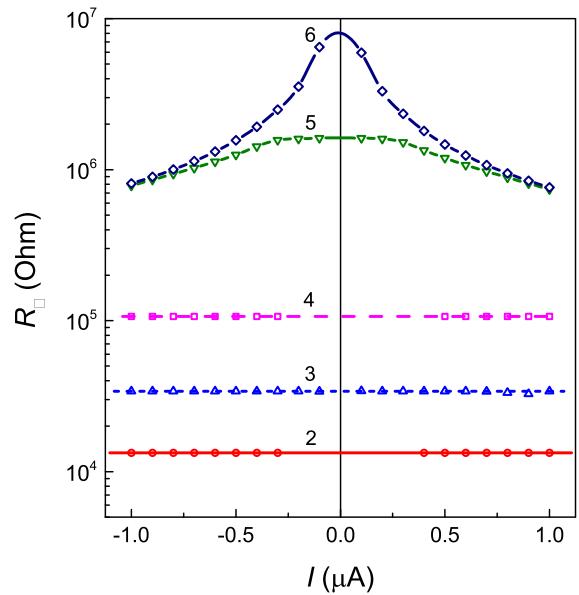}
\caption{Static resistance $R = V/I$ at $T=300$ K  as a function of current $I$
for samples 2 - 6.}
\end{figure}

We conclude with the statement that at the ultimate stage of disorder induced by ion irradiation, the graphene film ceases to be continuous and is fragmented to separated small-size graphene islands (quantum dots). Fragmentation of graphene film is accompanied by complete disappearance of the Raman spectra lines and change in the $I-V$ characteristics which reflects the different mechanism of conductivity. The maximal resistance of the continuous disordered graphene film is of order of reciprocal value of the minimal graphene conductivity.

\end{document}